\documentclass[11pt]{article}
\usepackage{moriond,epsfig}

\def\be{\begin{equation}}
\def\ee{\end{equation}}
\def\bea{\begin{eqnarray}}
\def\eea{\end{eqnarray}}

\begin{document}
\title{Unitarity of exclusive quark combination model: Exotic hadron production,
entropy change and charmonium production for  colour-singlet many-quark system}

\author{LI Shi-Yuan}

\address{School of Physics, Shandong University, Jinan, 250100, P. R. China}

\maketitle
\abstracts{
Confinement indicates an  asymptotic quark state  not observable except its energy is zero.
Unitarity indicates  that the total probability of a definite state of quark system
to transit to any final state  is exactly one.  This talk reviews some important
conclusions/predictions  from the basic properties like unitarity of the combination model,
as addressed by the title.}

\section{Introduction: Unitarity of exclusive quark combination model}

Quark Combination Model (QCM)  was proposed in early seventies of  20th century
(Anisovich, Bjorken) to describe the multi-production
process in high energy collisions, based on the constituent quark model of
hadrons. 
Various versions of     QCM  have  succeeded
in explaining many data. 
Recently in central gold-gold collisions
at the Relativistic Heavy Ion Collider (RHIC), 
 several `unexpected' phenomena  which
 lay difficulties for other hadronization mechanisms   
 can be easily understood in quark combination mechanism.
Common of all the hadronization models,
QCM responds to describe
the non-perturbative QCD phase. It includes two steps:  1) the
`partons' in various collisions turn into constituent quarks;   2) these
quarks combined into hadrons according to certain rules.
One can regard the combination model as a `reverse employment' of the
constituent quark model. 
In the following of this paper we 
concentrate on   step 2,
the  combination process, which is
the `realization' of confinement for the constituent quarks.
We will investigate the most general principles which a QCM has to respect, so
 that to see what can be reliablly {\it predicted}  by  such a model, rather than
 seek how to employ a certain version of QCM to make a
 good postdiction and parameterization of the data.
For this purpose, we deal with
a  colour-singlet (CS) system of many quarks prepared from step 1, but
 without addressing  how  the hard partons turned into constituent quarks, especially,
 `where is the gluon'?
(Prof. Dixon asked after this presentation) in step 1.
Charm and bottom
quarks are  produced form hard interactions. They in step 1
are `dressed' to be a constituent but their momentum spectra are not largely modified.
This special advantage will be discussed in the following. 

Without digging into details of any special kind of QCM,
one  easily figures out  two principles which it
 must respect:
 First of all, energy-momentum conservation is the
  principle law of physics,  reflecting   the basic symmetry  space-time displacement invariance.
  The models must precisely (as precisely as possible, in practice)
transfer the energy and momentum of the parton system into the constituent quark system
and then the hadron system.
Second, 
when applying the combination rules on  a CS separated system of
 constituent quarks, it is necessary that all the  quarks are
combined into hadrons,
 or else there are  free quarks with
non-zero mass and energy, which obviously contradicts to any
observations that suggest confinement.  Moreover, these free quarks take
away energy and momentum, hence make danger of the
energy-momentum conservation.  This second principle is referred as
Unitarity of the relevant model.
These two principles are closely connected, with the
first one 
the natural result of the second one.


  For a  QCM which respects and can reflect unitarity,
  the combination process can be described by a
unitary time-evolution operator $U$   , with
\begin{equation}
\label{ueq1}
\sum_h |<h|U|q>|^2=<q|U^+U|q>=1.
\end{equation}
 The quark state
 $|q>$,  describes a CS quark system,
  and the corresponding hadron state $|h>$ describes  the hadron system. The
matrix element $U_{hq}=<h|U|q>$ gives the transition amplitude.
For a separated system,
the energy-momentum conservation is inherent, by the natural commutation
 condition  $[U,H]=[U,\bf P]=0$, with $H, \bf P$ the energy and
momentum operator of the systems. This is just the
confinement which says that the total probability for the CS quark system
to transit to all kinds of hadron is exactly 1,
and agrees with the fact that all the constituent quark states and the hadron
states are respectively two complete sets of bases of the same   Hilbert space
\footnote{This is very natural,  if one adopts that  QCD is really the uniquely
correct theory for the hadron physics, with its effective Hamiltonian $H_{QCD}$.
Then all the hadron states with definite energy-momentum should be its eigenstates and expand the  Hilbert space of states (though we do not know how to solve $H_{QCD}$ mathematically).
 While  a model  is proposed  in language of constituent
quarks which composite the hadrons, all of the quark states with definite energy-momentum should be eigenstates of the same $H_{QCD}$
(Here we consider constituent quark model, and ignore the
rare probability of exotic hadrons like glueball, hybrid, hence need not consider gluon states).
So these two sets of bases are of different {\it representations},
as is more easy to be recognized if one imagines that  all
the wave functions of hadrons  are written  in terms of
 quark states in some special framework of quark models and
 notices that the planer wave function 
  as well as other
special functions (bound state wave functions) are all possible to be complete bases
 for a definite functional space, mathematically. },
i.e.,  $\sum |q><q|= \sum |h><h|=1$ for the colour-singlet system.
So combination process {\it never} changes the degree of freedom of the system.

In the following sections  we will address three relevant topics:
Unitarity of the  combination model
naturally suppresses the production of exotic hadrons; 
 Unitarity in exclusive QCM guarantees the non-decreasing of entropy
in the combination process for a CS separated system;
Unitarity does not introduce any new rules, when considering heavy quarks in
the combination. Sine lack of space, the Refs. are to be found from Ref. \cite{ind}.

\section{Exotic hadron (multi-quark states) production}

Two important points should  be considered:

1. As a matter of  fact from experiments,
\begin{equation}
\label{ueqless1} \sum_{h=B,\bar B,M} |<h|U|q>|^2 \sim 1-\varepsilon,
~ \varepsilon \rightarrow 0^+,
\end{equation}
here $B, \bar B, M$ denote baryon, antibaryon and meson respectively.
Na\"{i}vely from the group theory, {\it colour
confinement} seems not so restrict as  Eq. (\ref{ueqless1}). The CS
state, i.e., the invariant, totally antisymmetric representation of
the $SU_C(3)$ group, requires at least one quark and one antiquark,
or three (anti)quarks, but more (anti)quarks
can also construct this
 representation, hence possibly to form a CS ``hadron''. 
They are to be called exotic hadrons  (here not including glueball or hybrid).
 Until now, no experiment can definitely show    $\varepsilon$ in
Eq. (\ref{ueqless1}) is exactly 0 or a small but {\it non}-vanishing
number. If definitely $\varepsilon=0$, there must be underlying
properties of QCD which we still not very familiar. Even
$\varepsilon$ is not vanishing, its smallness, definitely confirmed
by experiments and shown in Eq. (\ref{ueqless1}),  also provides
interesting challenges, especially on hadronization models. The
small production rate of a special kind of exotic hadron seems easy
to be adopted. However, taking into account so many possibilities to
construct the CS representations by {\it various} numbers of
(anti)quarks, that the total sum of them is still quite small, is
very nontrivial as a property of QCD and even nontrivial for a
hadronization model to reproduce.

2. Colour recombination destroys the distinction between multiquark
state and molecule state.
All kinds of Exotic hadrons have one common property:
The (anti)quarks can be grouped into several clusters, with each
cluster {\it possibly} in CS. 
However,
the ways of grouping
them into clusters  are not unique, as
it is simply known from group
theory that the reduction ways for a direct product of several
representations are not unique.  Furthermore, these clusters need
not necessarily be in CS respectively, since the only requirement is the
whole set of these clusters in CS. For example, the system $q_1
\bar{q_2} q_3 \bar{q_4}$ (the constituents of a ``tetraquark'') can
be decomposed/clustered in the following ways:
$(q_1 q_3)_{\bar 3} \otimes (\bar{q_2}
\bar{q_4})_{3} \rightarrow 1,  ~~~
(q_1 \bar q_2)_{1~ or ~8} \otimes (q_3
\bar q_4)_{1~ or ~8} \rightarrow 1, ~~~
\cdot \cdot \cdot $
In the above example, only  the second case, when these two $q\bar
q$ pairs  are in CS respectively, it seems possible to be considered
as a hadron molecule. But dynamically, the colour interactions in
the system via exchanging gluons can change the colour state of each
separate cluster, so each kind of grouping/reduction way seems no special physical
meaning. Such an ambiguity, which has been considered in many
 hadronization and decay processes as ``colour
recombination/rearrangement'' 
obstacles the possibility
to consider the exotic hadron in a unique and uniform way, while
leads to the possibility of introducing some phenomenological
duality. Namely, even we consider the production of exotic hadron as
``hadron molecule'' formation, the subsequent colour interactions
 in the system can eventually transit  this
``molecule'' into a ``real''  exotic hadron, at least by some
probability.

From the above discussion, and  in the calculation by Shandong QCM (SDQCM),
 one can introduce a model dependent definition of
multiquark state, i.e., the number of quarks to be combined into the hadron
is definite though quark pair could be created in the bound state.
The fact $\epsilon \to 0^+$ is employed by introducing the parameter $x$.
It is clear that to an extreme if we have infinite
kinds of exotic hadrons, $x$ should be vanishing, expecting infinite
number of vanishing variables (production rates corresponding to  each  certain
 exotic hadron) summing up to get a finite small
result (the total production rate of all exotic hadrons).
So it is predicted that
if the Gell-mann Zweig quark model can be extrapolated to multiquark states, production
 of each of the species could be
vanishing and not observable.

\section{Entropy change}
1. By the formula of entropy $S=-tr(\rho \ln \rho)$
for a separated system,
we can conclude a unitary transition will not change the entropy.
%
\begin{eqnarray}
\rho(t)&=&|t, i>P_i<i,t|=U(t,0) |0,i>P_i<i,0|U^\dag (t,0) \nonumber \\
   &=& U(t,0) \rho (0) U^\dag (t,0).
\end{eqnarray}
 Here $U(t,0)$ is the  time evolution operator. $P_i$ is the probability of the
 state with index $i$.
Taking $\rho(0)$ as the distribution of the constituent quark system {\it just
 before combination}, while  $\rho(t)$ {\it just after}, of the hadron system,
then $U(t,0)$ is exactly the operator $U$ introduced in Eq. (\ref{ueq1}).
This is  a uniform unitary
transformation
on the Hermitian operator $\rho$, which  does not change the
 trace of $\rho \ln \rho$.
So  
the entropy holds as a constant in the combination process, same as energy and momentum.

2. Energy conservation is kept for each combination step  for the many quark CS separated system,
by tuning the constituent quark masses in the programme.  Then an  ideal quasistatic process can be employed
to calculate the entropy change. The result is again zero. The details are described in arXiv:1005.4664.

\section{Charmonium production}
In several combination models, including the SDQCM mentioned above,  one has considered the open charm
hadron production by introducing  the charm quark into
the bulk of the light quarks with its specific spectrum, to let all these four kinds of quarks to combine on equal footing. In this consideration, one has to deal with
 the case when a charm quark antiquark pair can be combined together under the
 combination rule to keep consistency. On the other hand,
one can raise the question whether charmonium (or bottomium) can be produced under exactly
the same mechanism as light $q \bar{q}$ hadron, i.e.,
via the  common combination rules.

Charm/bottom is the kind of 'constituent quarks' which is more easy to be tracked
than the light ones,
and the `dressing process'  will not change the
spectrum much. The light quark sector,  many of them come from gluon nonperturbative QCD transition,
which is yet quite unclear, as described above.
So it is  more reliable talking about the charm distribution before combination.
The above investigation of charmonium as well as the open charm in QCM can help
the study of its energy loss in medium.

When according to the combination rules, a $c \bar{c}$ pair can be combined,
we further restrict their invariant mass to lower than some definite value (say, that of  $\Psi(3S)$)
to be a charmonium.  This will not change the unitarity mentioned above.
This has a good analogy:  the charmonium and open charm corresponding to the positronium and free electron
(discrete and continuous spectrum), respectively.
Such a restriction does not affect comparing with data, either, since charmonium resonances more massive than
$\Psi(3S)$ almost all decay into open charms.

Our results indicate that at RHIC, the charmonium can be described exactly in the same way of
the open charm particle by SDQCM without introducing any new rule.
This check is to be done for LHC soon. 

More details in the long write up will come soon.
A preliminary figure can be seen from  the presentation slides, P9;
A table and formulae  show the relative ratio of different kind of charmonium, see P8.

\section{Postscript}
In the above section, when comparing our result on $J/\Psi$ spectrum and concluding
 consistent with RHIC data,
we neglect the contribution of bottom.
In  higher energies and larger transverse momentum $P_T$,
e.g., in  LHC, the contribution of B decay will increase and could be dominant for enough large $P_T$. In this case one need a separation of prompt in experiments, as done by CDF in Tevatron.
 To coincide with inclusive data, the theoretical calculation must include the bottom production.
 On the other hand, $J/\Psi$ is a good measure of B for large $P_T$ (Two body decay to $J/\Psi+h$ is an important way to measure B).
Such a fact can be seen from the talks in this Rencontre (Z. Dolezal, K. Ulmer).
Combined with the celebrating $J/\Psi$ suppression data in Pb-Pb of large $p_T$
 reported  by ATLAS in this Rencontre (B. Wosiek),
it is easy to conclude that the bottom energy loss is almost the same as the
light quark for large $P_T$, as expected by the author in discussions around  the dinner table in La Thuile.
The 'non-photonic' electron and forward muon data measured by ALICE presented on QM2011 (Annecy, May) is not contradicted with such a expectation, though the $R_{AA}$ a little larger comparing to e.g., pion. However, one sees $R_{AA}$ increasing with $P_T$ and the $P_T$ of electron/moun represents the behaviour of B around $2P_T$.
Such a energy loss behaviour is well understood by considering the spacetime picture of
the jet (heavy or light) medium interaction, as explained in the talk by the author in last year's
Rencontre. This kind of interaction has an analogy of hadron hadron interaction. The production process (pionization)
is the main mechanism to lose energy. The produced particles composite the rapidity plateau, so that can be of large angle w.r.t the jet.
Since the width and height of the plateau increase with the interaction energy, the energy loss also increases
with jet energy.
These have been confirmed by CMS measurements reported in La Thuile (F. Ma). 
$\frac{\Delta E}{E}$ is a constant for a large range of jet energy.



\section*{Acknowledgments}
 My collaborators (HAN W., JIN Y., SHANG Y., SHAO F., SI Z., YAO T., YIN F.)
in doing the works presented here are first of all
thanked.  The works are partially supported by
NSFC with grant Nos. 10775090, 10935012,  NSFSC with grant Nos. ZR2009AM001, ZR2010AM023, JQ200902,
and Independent Innovation Foundation of SDU.

\end{document}